\documentclass[12pt,sans]{article}
\usepackage{amsmath}
\usepackage{amsfonts}
\usepackage{amsthm}
\usepackage{amssymb}
\usepackage[sans]{dsfont}
\usepackage[final]{graphicx}
\usepackage{mathrsfs}
\usepackage{nicefrac}
\usepackage{sans}
\usepackage{siunitx}
\usepackage{bm}
\usepackage{color}

\usepackage[T1]{fontenc}
\usepackage[ansinew]{inputenc}

\begin{document}

\begin{center}
{\large {\bf Uncertainty in Climate Science: Not Cause for Inaction}}\footnote{This article appeared in print in two parts (without the material in the appendix of the present article), covering separately the two themes of this article: J.M. Restrepo, M. Mann, {\it This is how Climate is Always Changing}, SIAM News May 2018, and J.M. Restrepo, M. Mann, {\it This is how Climate is Always Changing}, Focus Group on Climate
Newsletter, American Physical Society, March 2018.  For citation purposes please use both of these articles. If a technical aspect presented in the appendix of this paper are required, please add a citation to this article as well.}

Juan M. Restrepo, \\
Computer Science and Mathematics Division,
Oak Ridge National Laboratory, \\
Oak Ridge TN, USA 37831\\
Michael E. Mann,  \\
Department of Meteorology \& Atmospheric Science,
and The Earth and Environmental Systems Institute,
 The Pennsylvania State University
 University Park, PA 16802
 \end{center}

The Fourth National Assessment, Climate Science Special Report of the US Global
Change Program,  published in November 2017,  concludes,
``based on extensive evidence, that it is extremely likely that human activities, especially
emissions of greenhouse gases, are the dominant cause of the observed warming since
the mid-20th century." 

When asked about some of the conclusions in the report regarding systematic climate
change, Mr. Raj Shah, a spokesman for the Trump administration, stated, "The climate
has changed and it is always changing. "
 Shah is echoing assertions from  other observers that there is nothing unusual
about the changes in climate and weather that we are experiencing: There have
been changes before the industrial era, and some of these have been extreme.
 Translated
to more technical terms, such observers claim that climate has a stationary statistical distribution --one that does not change with time-- and  that, in recent
years, we just happen to be experiencing samples of this distribution that are, possibly rare,  extreme
highs. 
Shah continues, "[As the report] states, the magnitude of future
climate change depends significantly on remaining uncertainty in the sensitivity of Earth's
climate to greenhouse gas emissions"   \cite{eos17}. 
 This part is  hard to interpret, but it is meant to imply  that climate forecasting is made unreliable by the presence of uncertainties. 
 Shah's statement  is fully consistent with the US Environmental Protection Agency's (EPA) Pruitt's statements, such as 
``I think that measuring with precision human activity on the climate is something very challenging to do and there's tremendous disagreement about the degree of impact," he told CNBC.
``So no, I would not agree that it's a primary contributor to the global warming that we see," Pruitt said. ``But we don't know that yet, we need to continue to debate, continue the review and analysis''  \cite{reuters}.
Acknowledging that climate changes but that the anthropogenic contribution is, {\it at the same time}, too difficult to estimate and too small to be of importance, is a frequently-used assertion of  the Trump administration. Rather than disentangling the statement, we will instead address separately the issue of statistical stationarity of climate time series, and how climate predictions are impacted by uncertainties in natural and anthropogenic forcings. 

 To explore the assertion of  a static climate distribution, we introduce here a theorem that applies to record values of
a series of random variables drawn from a stationary distribution, such as the measured temperatures. 
Others have used this approach more rigorously to examine trends in climate data \cite{benestad,krug07,wergen13}.
Since the only requirement made in the theorem on the random variables is that they derive from a stationary distribution, the failure of this theorem to hold indicates that the distribution from which this data arises is not stationary.  The theorem or its application does not yield  causal attributions to its outcomes. 
Nevertheless,  the use of such a simple test circumvents the necessity to argue about data statistics based upon model outcomes.  

We will also examine how the inclusion of historically-informed uncertainties on natural and anthropogenic greenhouse gases (GHG) 
modify climate predictions. To do this we  will use a simple model that captures  the essential phenomenology of the  radiation balance of 
more complete state-of-the-art climate models. This {\it energy balance model} (EBM) will be used in what follows to determine whether uncertainties in GHG emissions  lead to qualitatively different climate projections than those obtained without taking uncertainties into account.

Before proceeding we should clarify  what is meant by  climate, as opposed  to
weather. Climate and weather describe the same system, but the term 'climate'
refers to large spatio-temporal scales and 'weather' to small ones. This distinction
is not just a convenience. While both describe the energetics, mass and momentum
exchanges of a rotating Earth, the scale determines the prominence of the different phenomenology, weather
being dominated by inertial effects (advective processes, turbulence, waves, density dynamics, and transient
and sometimes unstable conditions) and climate by the forced/dissipative effects (radiation
and ocean and atmospheric transport). Weather includes tornados, hurricanes,
or extreme values of temperature or rainfall. Examples of climate are the seasons, El
Ni{\~n}o/Southern Oscillation, the ice ages and  Industrial Era global warming.

\section*{Records in Time Series}
 
 We make  the following assertion: climate temperatures are samples from a stationary distribution.  If so, a theorem that applies to stationary distributions should be borne out by the data. We apply a theorem about record highs and record lows (see \cite{fosterstuart}).

One draws a sequence of independent and identically-distributed (IID) samples $X_1, X_2, \ldots$  from a stationary distribution. We denote  a sample from the sequence
a {\it record} high (low) if its value is higher (or lower) than the samples preceding it.
The 
probability of a record high is $P_n := \mbox{Prob}[X_n > \max\{ X_1, X_2, \ldots, X_{(n-1)} \}]$ (with the obvious modifications for the record low). 
In a sample set of size $n$ any one particular value has equal chance of being the greatest (lowest) value, thus $P_n = 1/n$.
We denote  as $\mathbb{E}(R)$  the expected number of records for a stationary random sequence 
 of size $n$. It  is given by the harmonic series
$\mathbb{E}(R) = 1 + 1/2 + 1/3 + \cdots + 1/n $. For large $n$, $\mathbb{E}(R) = \gamma + \log(n)$, where $\gamma$ is the Euler constant. 

The occurrence of record values in climate data has been carefully compared to predictions for a stationary distribution (see, for example, 
\cite{benestad}, \cite{krug07}, \cite{wergen13}).
  Just to convey a feeling for such analyses, we undertake here a much less rigorous but hopefully illuminating look at some data. It should be cautioned that the application of this theorem to real data is highly nontrivial. Hence, in what follows, we will be using this exercise merely to give a suggestive outcome.
  
If the theorem  applies to climate data, we expect to wait increasingly long times for each new record temperature value (either high or low) because the probability declines as $1/(t-t_0)$, where the time $t$ of each temperature observation takes the place of the statistical index $n$, and $t_0$ is the start of the particular temperature observations.   (If the probability distribution  were symmetric we would also
expect the rates of record highs and  lows to be similar).

\begin{figure}[h] \centering
(a)\includegraphics[height=2in,width=5in]{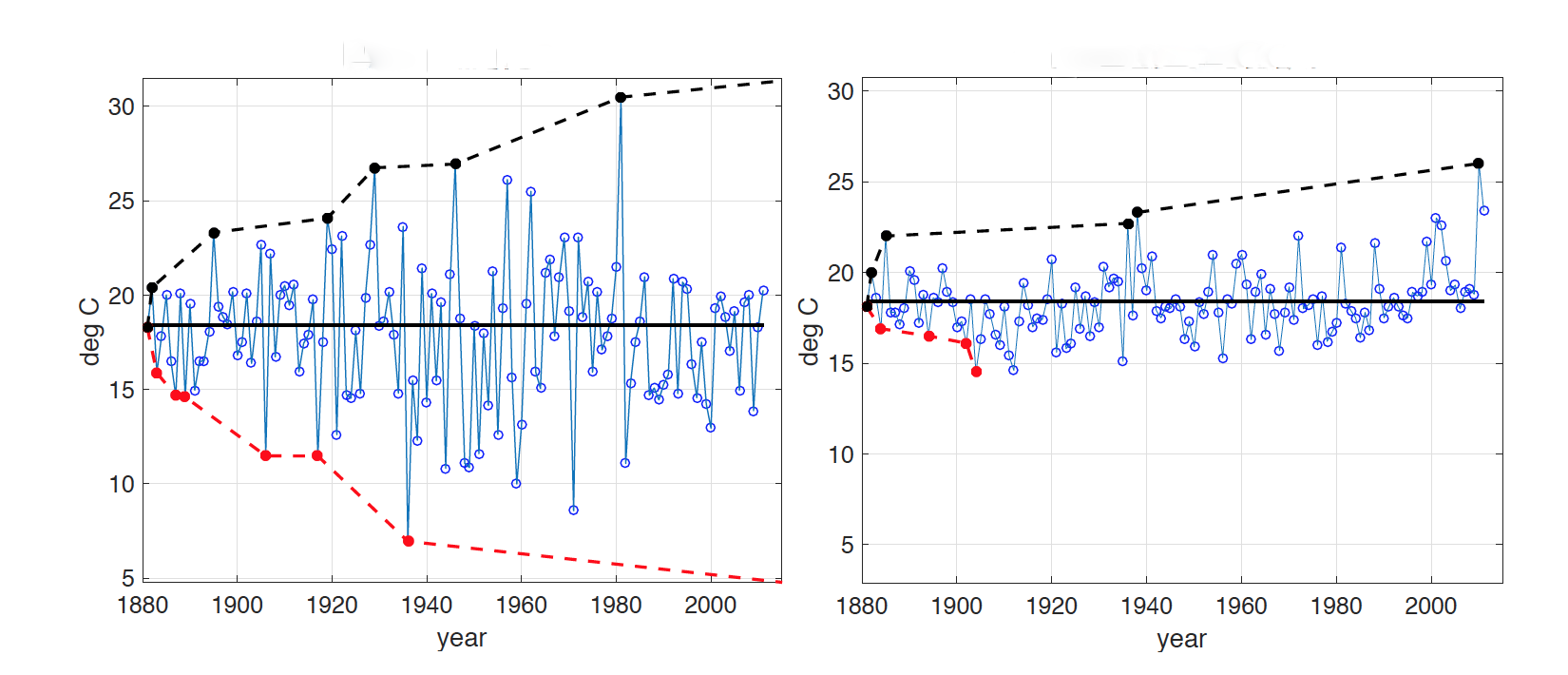}
(b)\includegraphics[height=2in]{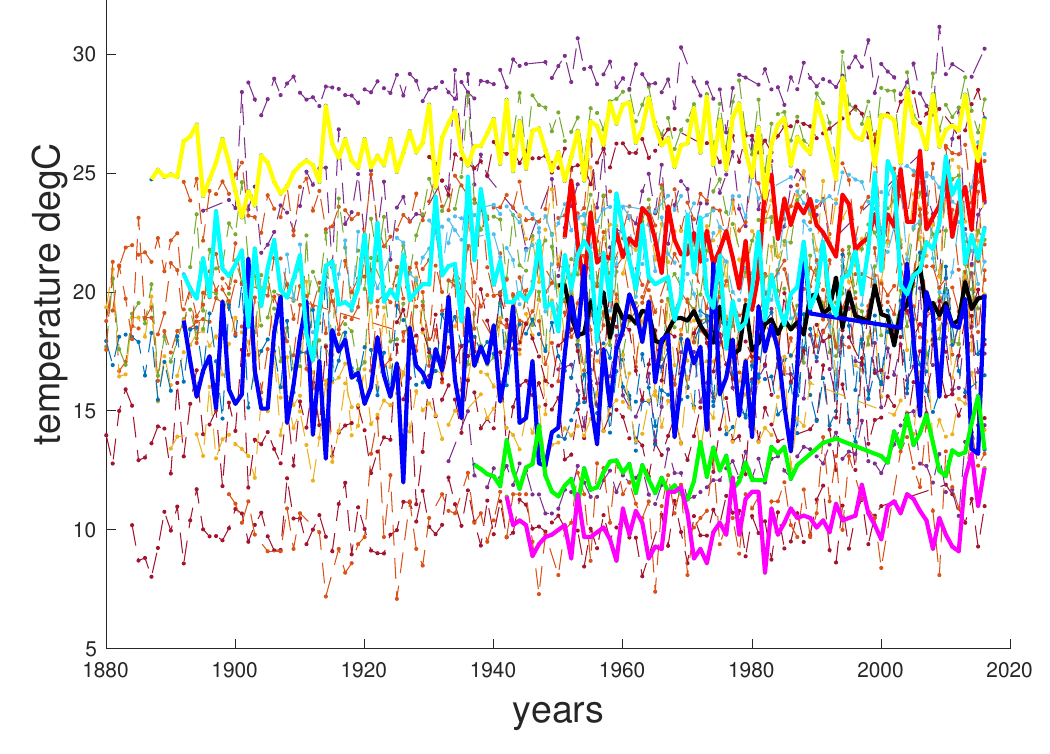}
(c)\includegraphics[height=2in]{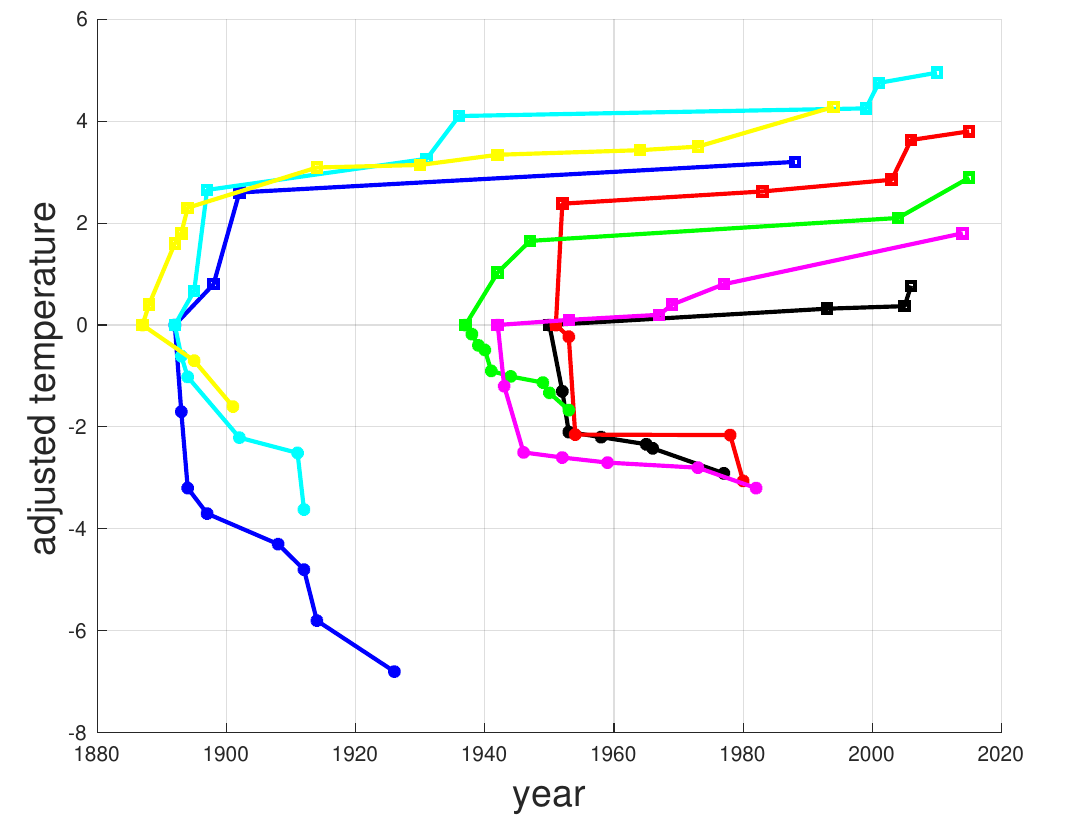}
	\caption{{\it (a) (left) records in a synthetic stationary distribution, (right) records of July
monthly temperatures at the Moscow station; (b) temperature data, as a function of
time, for 30, arbitrary locations in the Northern Hemisphere; (c) Record values for seven temperature time series,
highlighted in (b). The adjusted temperature subtracts out the first temperature value
in the time series. The data is taken from GISS repository. We note that you reach a time in each data set beyond which no new lows occur whereas new highs continue to appear going forward in time.  Temperature in degrees $^o$C.}}
	\label{fg.temps}
\end{figure}

Figure \ref{fg.temps}a compares the record high/lows obtained from a synthetic random time series to the July temperatures measured at the Moscow
station, from about 1880 to 2011 \cite{giss}.  For the random time series, the highs and lows are similarly spaced in time. However, for the Moscow temperature data, one sees many lows occurring at the early times and none after about 1910. By contrast, the record highs are more spaced out in time and continue through the observation period shown. The data suggests that the theorem on records is not fulfilled  and that the rate at which record highs  or lows occur at time t does not follow $1/(t-t_0)$. 

Figures   \ref{fg.temps}b and c  plot temperature data (from \cite{giss}) from 30 locations in the Northern
Hemisphere. 
The locations were chosen at random but were mostly concentrated around
temperate zones, simply because these records tended to be longer. The time series are not
all the same length and some stations did not report every year. 
The superposition of the data in Figure  \ref{fg.temps}b would lead you to believe that, over the course of the industrial revolution, a stationary distribution of temperatures is not all that bad a statistical model.
 In that figure we highlight seven temperature time series, chosen arbitrarily. The records associated with these 7 data are plotted in Figure \ref{fg.temps}c. To facilitate comparison, these seven data sets have been adjusted by subtracting the first temperature in the set (thus the adjusted temperature of any of these time series was 0). 
 Adding more observations to the top set or more  observations to the bottom
set does not change the impression that, with time, more high records will occur than low records
(the low records stop occurring). The key observation is that the record highs and the
record lows do not obey the $1/t$ dependence. Another test would be to compute the
expectation of the number of records to see if  it is logarithmic, but  doing so requires either longer data sets or a larger collection of data sets, such as the combined readings of all stations in the US. 

The key observation is that the record highs and the
record lows do not obey the $1/t$ dependence. Hence,  these temperature records  must not samples from a stationary
process.

Wergen and Krug  \cite{wergenkrug}, \cite{wergen11}, \cite{krug07} and others, have taken this line of research much further, to include correlations, a multiplicity of distributions, consideration of spatial dependence.  They have also found that removing a trend in the data makes the theorem 
more likely to be consistent with the statistics of temperature data.
The values obtained for the trend, using this analysis, are consistent with estimates of a rate of increase in the global mean temperature of
about 0.7$^o$C  in the land/ocean temperature over the last century, roughly ten times faster than the average rate of ice-age-recovery warming (see NOAA web site. Also see https://globalclimate.ucr.edu/resources.html for educational material on this topic).

In summary, the data suggests that Earth's climate is a non-stationary process. This
is something climate scientists would find consistent with   what we know
about climate.  Hence, it is not likely that the temperature extremes that
we experience today are rare events, but rather, the result of a changing climate. 
The findings of this line of inquiry, using much more technical assumptions and allowing for
correlations and for a multiplicity of probability distributions, indicates that climate has
been severely biased upward during the Industrial Era. The use of this theorem to estimate the return time of record high temperatures
would seriously underestimate the occurrence of historical highs, and overestimate
historical lows.

\section*{Incorporating Uncertainties in Greenhouse Gas (GHG) Projections}

Global estimates of GHG emissions are readily available (see \cite{co2estimate}) and their estimates have tightly constrained uncertainties, since these are critical to the economy of  the energy sector. Private and public entities keep track of production and resources fairly accurately. 
Uncertainties are associated, however,  with how to regulate and tax GHG. This uncertainty is financially significant, but these
would  affect the design of commodity trading treaties, just as uncertainties in many other commodities routinely  affect how they are traded.
Uncertainties in the effect of GHG  on climate play out locally: there are well-recognized uncertainties related to transport and dispersion as well
as to critical dependencies on their interaction with their environment (land, oceans, atmosphere). However, 
we are going to focus here on variability spanning several decades to hundreds of years  and the largest of spatial scales. At these scales  local nuances 
are not resolved and a balance model yields the temporal evolution of a global temperature $T$.

By incorporating uncertainties into the temperature and radiative forcings we can explore to what extent uncertainties in GHG affect conclusions of future projections
of the temperature. The uncertainties are derived from a statistical analysis of the historical data. We can compare natural and anthropogenic GHG forcings, taking into account uncertainties, in order to determine whether the sensitivity of the outcomes depends on the relative uncertainties in these two GHG components. We can also infer whether natural or anthropogenic forcings are dominant, prior to the Industrial Era, during this Era, and in the future.

Black body radiation tells us that the earth's radiation is  proportional to $T^4$.  The surface energy balance, in terms of the surface temperature $T$,  is $C dT/dt = Q + \kappa  \sigma T_{Atm}^4 - \sigma T^{4}$, where $T_{Atm}$ is the atmospheric temperature, $t$ is time, $C$ is the effective heat capacity, and  $\sigma$ is Stefan-Boltzmann constant.  $Q$ is the effective incoming radiation. 
 If $C_a dT_{Atm}/dt$ is small, where $C_a$ is the atmospheric effective heat capacity, then
$\kappa \sigma T^4 + 2 \kappa \sigma T^4_{Atm} \approx 0$, then $C dT/dt = Q -(1-\frac{\kappa}{2}) \sigma T^4$. The range of temperatures of the process is not large, hence, $(1-\frac{\kappa}{2}) \sigma T^4 = A + BT$, is a linearization and the constants in the 
nonlinear formula are subsumed by $A$ and $B$. The energy balance is spectrally dependent: the high frequency component has a portion that 
reflects back to space by clouds and snow/ice and one that mostly dissipates. The low frequency component, on the other hand, is affected by reflectivity and the presence of a complex layer of gas, dust, and droplets which is capable of trapping the surface outgoing  radiation. 
$Q$ will be assumed to be a linear combination of the effective solar radiation and the radiative forcing due to GHG. Hence $Q = \frac{1}{4}(1-\alpha) S + 
F_{GHG}$, where the albedo $\alpha \approx 0.3$, and  the global average  solar radiation $S/4 \approx 1370/4$ Wm$^{-2}$, presently. (See   \cite{mcguff}, \cite{ebmnorth}, and \cite{mann14}). The Energy Balance Model (EBM) we adopt is thus
\begin{equation}
C dT = \frac{S}{4}(1-\alpha) dt + F_{GHG} dt -(A+B T)dt  + \nu(t),  
\label{ebmeq}    
\end{equation}
where $T$ is the temperature of Earth's surface (approximated as the surface of a 70-meter-depth, mixed-layer ocean covering 70 percent of 
Earth's surface area).  $C = 2.08 \times 10^8$ J K$^{-1}$m$^{-2}$ is the effective heat capacity that accounts for the thermal inertia of the mixed-layer ocean, but does not allow for heat exchange with the deep ocean as in more elaborate 
'upwelling-diffusion models' (\cite{wigley90}). The last term in the equation is a stochastic forcing term,  added to represent inherent uncertainties and unresolved processes (Hasselmann, \cite{hass76}, spearheaded the conceptualization of low dimensional models
for climate with an unresolved physics that had its own, possibly stochastic, dynamic). The model (\ref{ebmeq}) appears in  \cite{sciammann} (and a link is provided to matlab code which can be used to reproduce the temperature  outcomes in that article).  The choice $A = 221.3$ Wm$^{-2}$ and $B = 1.25$ WK$^{-1}$m$^{-2}$ yields a realistic preindustrial global mean temperature $T = 14.8 ^o$C and an equilibrium climate sensitivity (ECS) of  3.0 $^o$K, consistent with midrange estimates by the International Panel on Climate Change (\cite{ipcc07}), as the range is likely somewhere between $1.5-4.5$.  The equilibrium climate sensitivity is defined as the temperature change due to the effect of doubling in the concentration of  CO$_2$ in Earth's atmosphere.  A $1^o$K change requires a doubling of CO$_2$, or equivalently,  to an increased
  forcing of 3.7 W m$^{-2}$.

The model was driven with estimated annual natural and anthropogenic forcing over the years A.D. 850 to 2012. 
Greenhouse radiative forcing was calculated using the approximation $F_{GHG} = 5.35 \log(\mbox{CO}_{2e} /280)$, where $280$ parts per million (ppm) is the pre-industrial CO$_2$ level and CO$_{2e}$ is the 'equivalent' anthropogenic CO$_2$  (see \cite{myhre98}). The CO$_2$ data from Ammann {\it et al} \cite{ammann} is used, scaled to give CO$_{2e}$ values 20\% larger than CO$_2$ alone (for example, in 2009 CO$_2$ was 380 ppm whereas CO$_{2e}$ was estimated at 455 ppm). 
Northern Hemisphere anthropogenic tropospheric aerosol forcing was  taken instead from Crowley 
  \cite{crowley}, with an increase in amplitude of 5\% to accommodate for a slightly larger indirect effect than in \cite{ammann}. A linear extrapolation of the original series (which ends in 1999) is used to extend though 2012\footnote{One needs to distinguish between emissions and CO$_2$ concentrations. The latter
are directly measurable in the atmosphere and the increase is monotonic. This forms the basis for the extrapolation model.}.
 In the simulations 
we have assumed that tropospheric aerosols decrease exponentially from their current values with a time constant of 60 years. This gives a net anthropogenic forcing change from 2000 to 2100 of 3.5 Wm$^{-2}$, roughly equivalent to the International Panel on Climate Change's 5th assessment report 'RCP6' scenario, a future emissions scenario that assumes only modest efforts at mitigation.
Estimated past changes in solar irradiance were prescribed as a change in the solar constant $S$ whereas forcing by volcanic aerosols was prescribed as a change in the surface albedo $\alpha$. Solar and volcanic forcing were taken from the General Circulation Model (GCM) simulations described in  McGuffie and Henderson-Sellers \cite{mcguff}. These were modified as follows:
the solar forcing was rescaled under the assumption of a 0.1 percent change from Maunder Minimum to present, more consistent with recent estimates (\cite{ammann});  the volcanic forcing applied was the mean of the latitudinally varying volcanic forcing in Ammann {\it et al} \cite{ammann}. With no  added uncertainties, solar output and no climatically-significant volcanic eruptions were assumed, for the years 2012-2100.
CO$_2$ radiative forcing has been extrapolated linearly, till the year 2100,
based on the trend over the past decade (which is roughly equivalent, from a radiative forcing standpoint, to a forward projection of the exponential historical trajectory of CO$_2$ emissions). A net tropospheric
anthropogenic aerosol  forcing rate decrease of 0.7 Wm$^{-2}$ per year, starting with the year 2000 has been taken into account in the model.



. 

\begin{figure}[h] \centering
\includegraphics[scale=0.7]{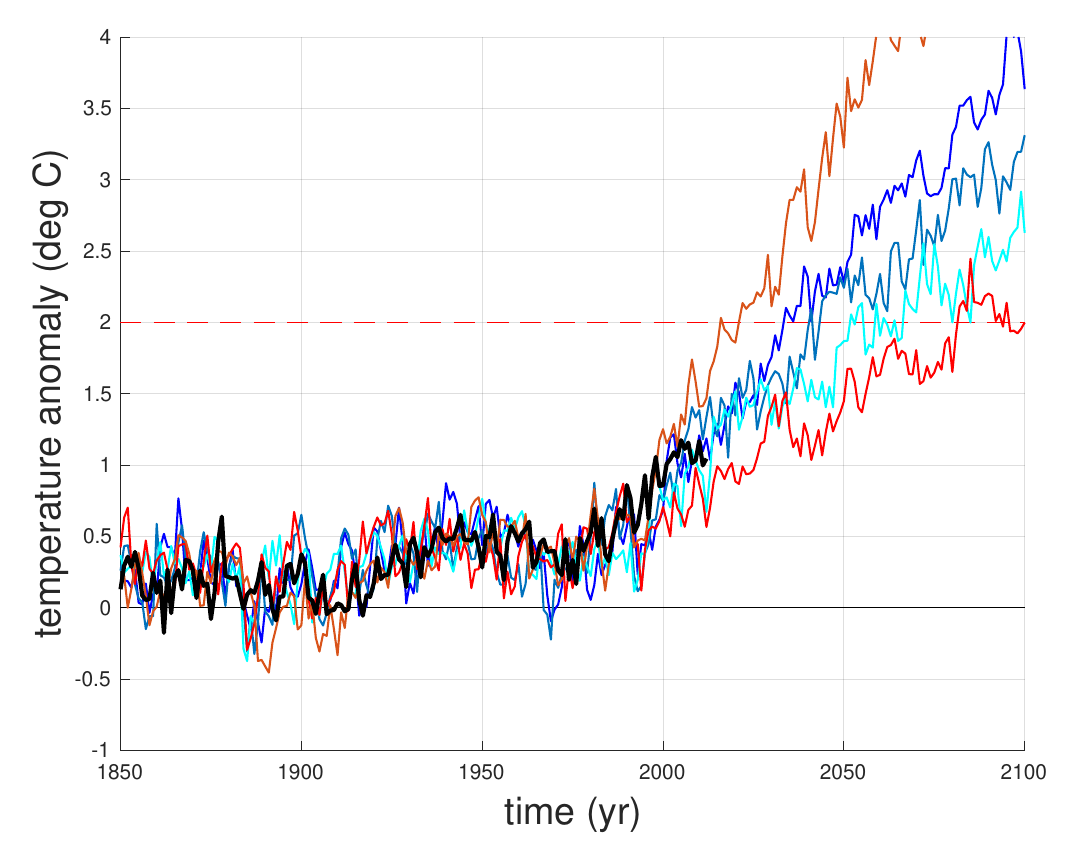}
	\caption{ Temperature prediction, as a function of time, for various values of 
	equivalent climate sensitivity (ECS). Shown, a single realization for each  ECS. The stochastic model for the temperature was informed by instrumental data. From top to bottom, $ECS=4.5,3,2.5,2,1.5$. The instrumental record appears in black.}
	\label{fg.steady}
\end{figure}

\begin{figure}[h] \centering
(a)\includegraphics[scale=0.6]{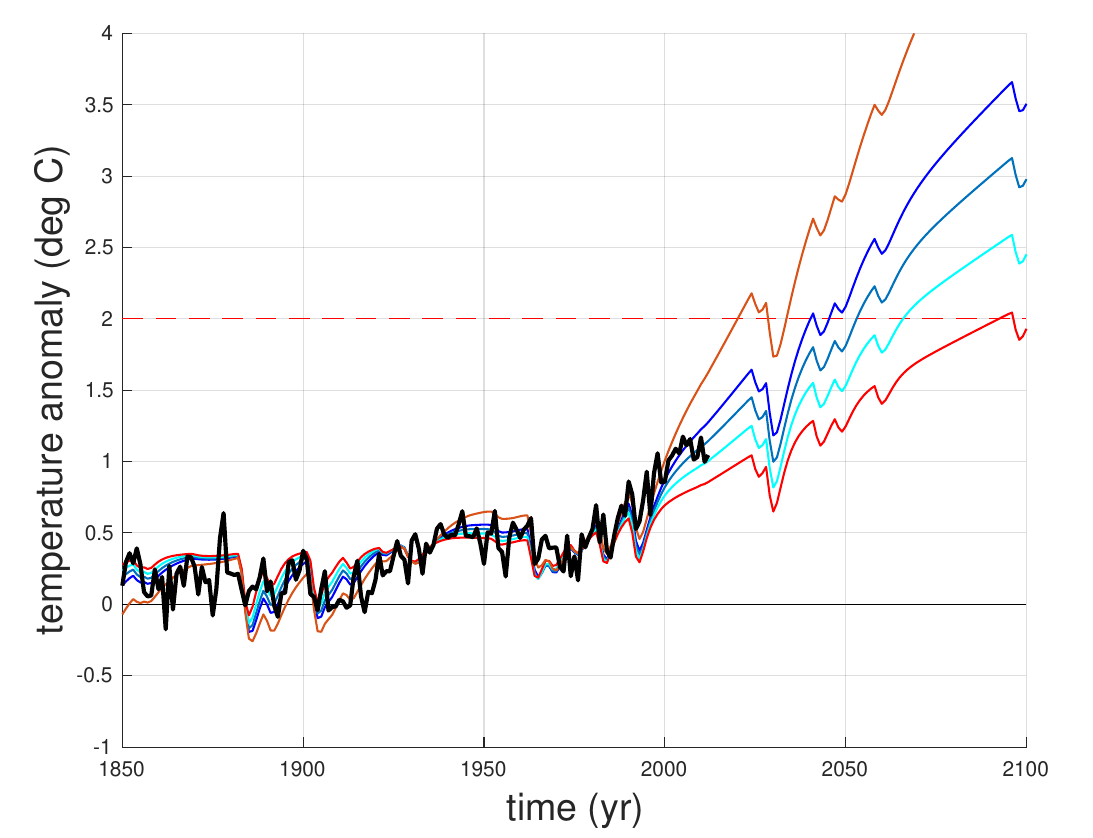}
(b)\includegraphics[scale=0.6]{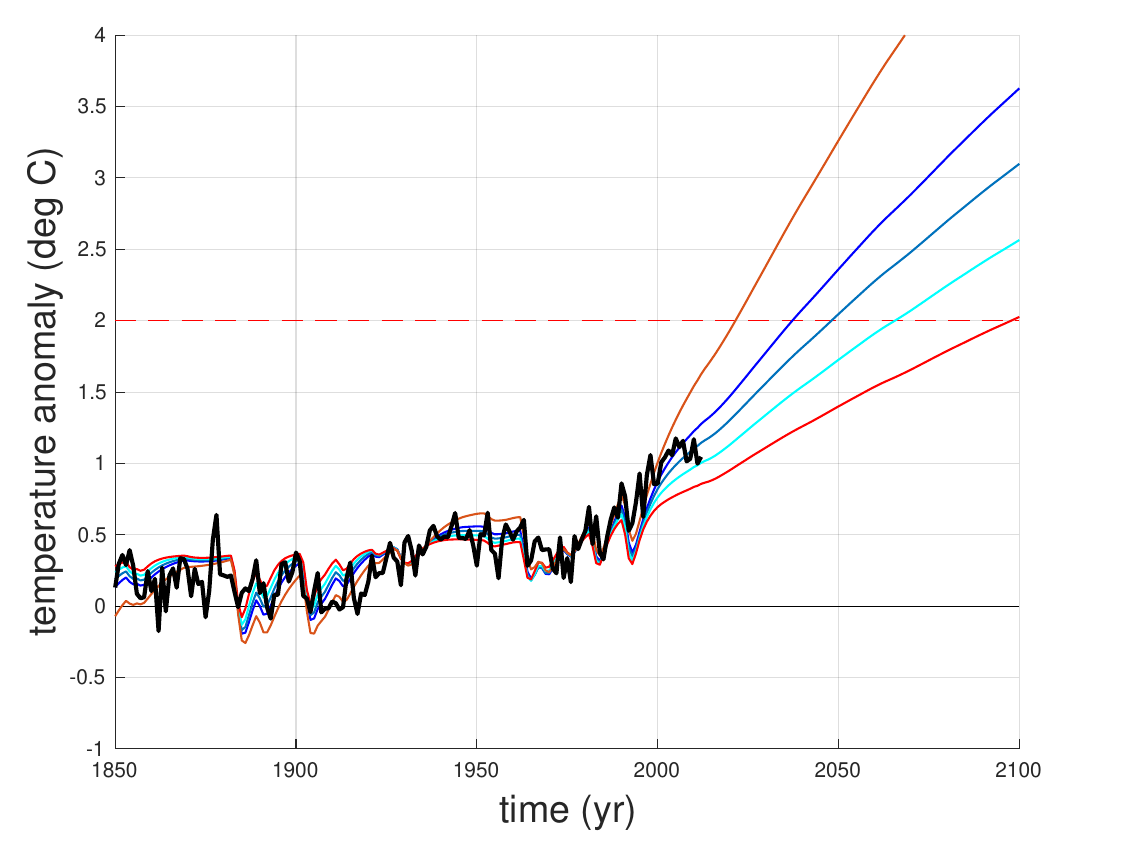}
	\caption{Temperature predictions, as a function of time for various equivalent ECS values.  A single realization is depicted, and the same stochastic model outcome is used in each of the ECS cases.
	In (a) predictions are made by accounting, exclusively,  for stochastic volcanic forcing variability. In (b)  for stochastic solar forcing variability exclusively.
	From top to bottom, $ECS=4.5,3,2.5,2,1.5$. The instrumental record is depicted in black.}
	\label{fg.volcsol}
\end{figure}

 As summarized in the  IPCC report  the net rise in temperature during the Industrial Era is due to  the upward trend of the anthropogenic GHG forcing. What's more, there is no way to predict the temperature increases measured during the Industrial Era, solely, by natural forcing. 
 Figure \ref{fg.steady} depicts EBM temperature predictions as a function of various values of the equivalent climate sensitivity (ECS) with 
 historically-based additive uncertainties added. Variations on ECS convey the sensitivity of the predictions to epistemic uncertainties in the models.
The instrumental record appears in black in the figure.
The temperature predictions shown represent  single realizations of the energy balance model. Details of the temperature noise model appear
in  the Appendix.  
The outcome suggests that taking into account historically-informed temperature variability in the predictions does not alter 
the conclusion that the upward trend in the anthropogenic forcing remains dominant factor in explaining temperature rises during the Industrial Era. 
We examine next whether taking into account stochastic variability in the natural as well as
 anthropogenic forcing changes this conclusion.
 Figure \ref{fg.volcsol} portrays temperature predictions from EBM model runs that incorporate uncertainties in the natural forcing.  In Figure \ref{fg.volcsol}a   we include, exclusively, volcanic forcing uncertainties. In Figure \ref{fg.volcsol}b we allow for 
 solar forcing uncertainties. The various curves result from  using different ECS values.   The description of the uncertainty models for volcanic, solar, and CO$_2$  source  variability is found  in 
 the Appendix.  Plotting a single realization, it is hoped, conveys qualitatively the relative impact of the stochasticity on 
 the outcomes (qualitative characteristics of the sample moments can be easily inferred from  the specific noise models used).
 As is expected, the temperature reacts abruptly  to highly localized volcanic emissions and
the sensitivity to solar variability is small, comparatively.  Hence, a significantly larger portion of the variance in an ensemble of these predictions is attributed to the volcanic emissions, rather than the solar uncertainties. 

\begin{figure}[h] \centering
\includegraphics[scale=0.7]{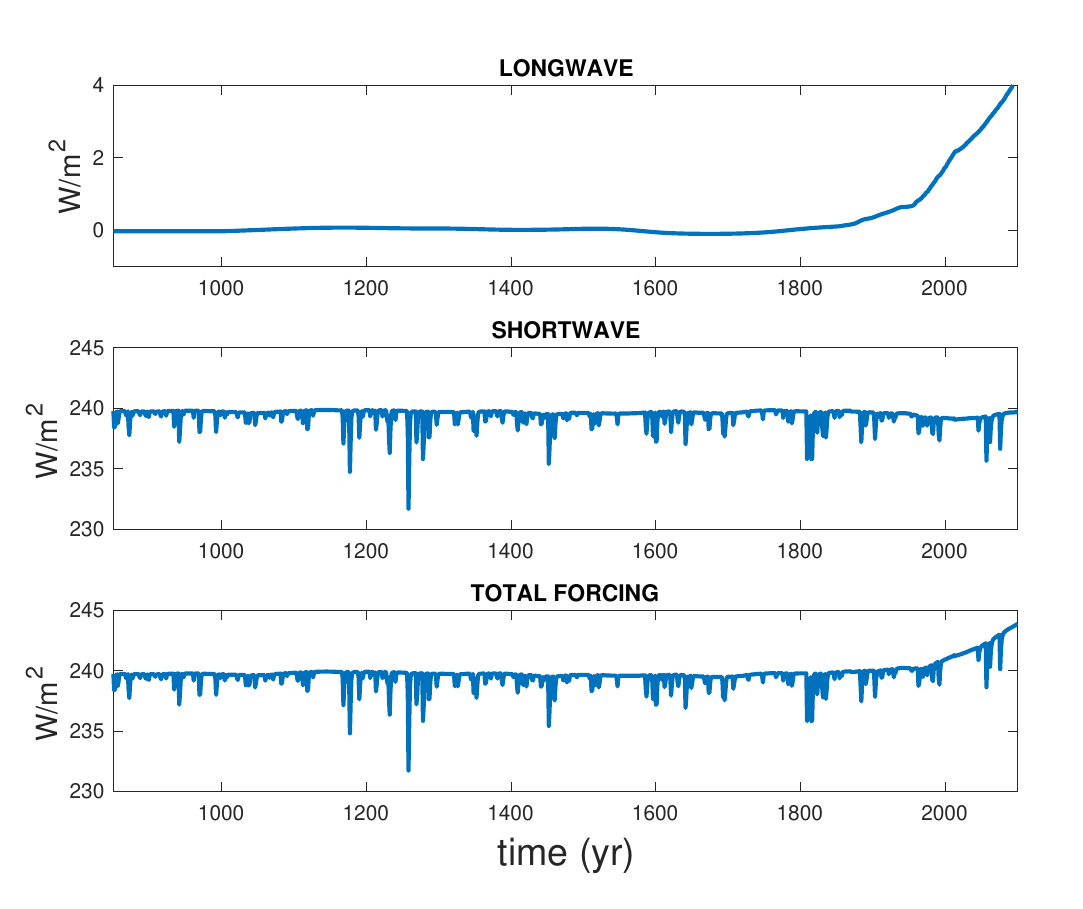}
	\caption{Stochastic long wave, short wave forcing and composite total forcing,  with uncertainties due to CO$_2$ emissions, volcanic activity, and solar forcing. A single stochastic realization is depicted. }
	\label{fg.forcings}
\end{figure}

\begin{figure}[h] \centering
(a)\includegraphics[scale=0.6]{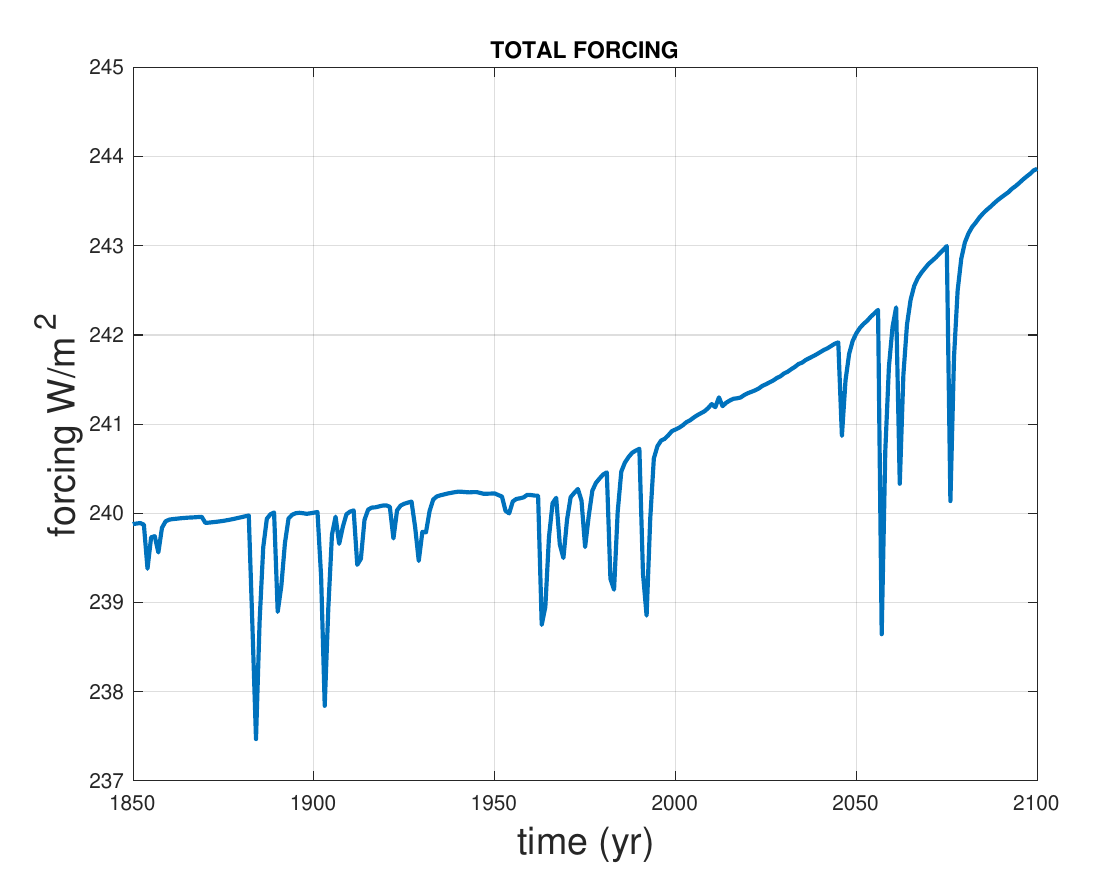}
(b)\includegraphics[scale=0.6]{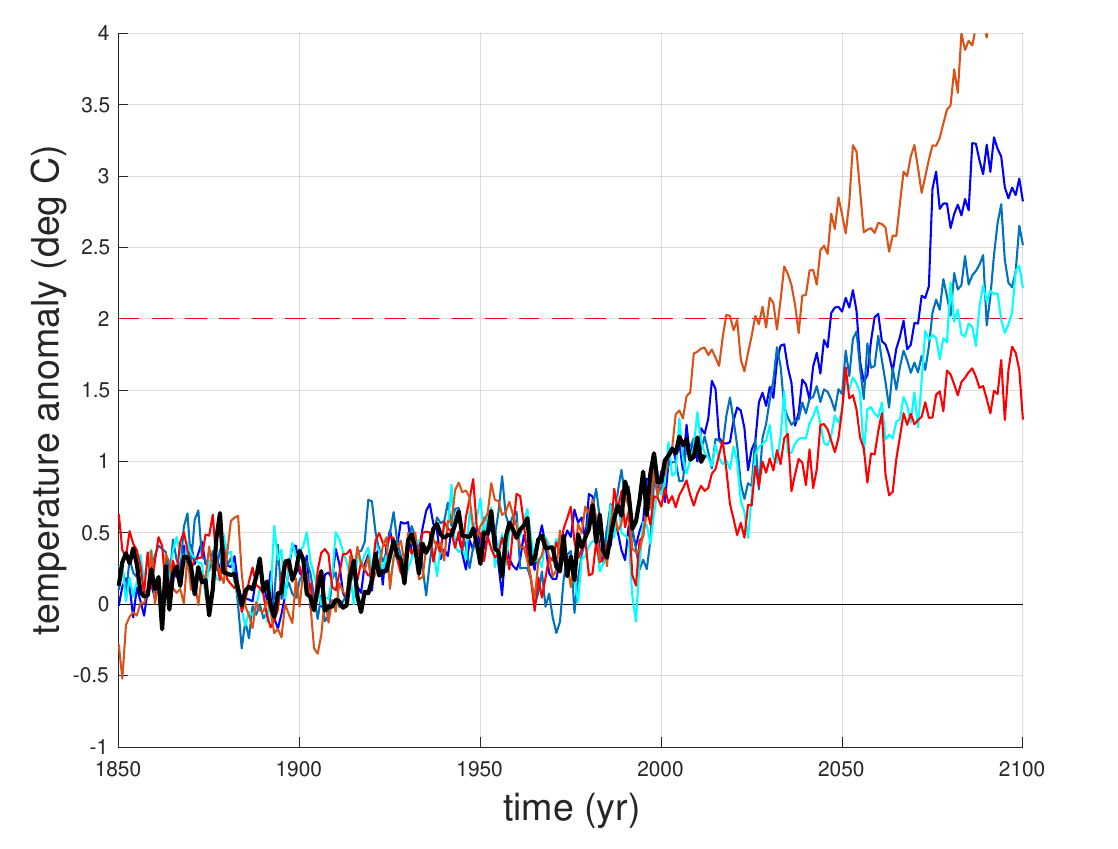}
	\caption{(a) Highlight of the composite forcing appearing in  Figure \ref{fg.forcings} corresponding to the period 1850-2100;  (b) temperature predictions as a function of ECS, taking into account  uncertainties due to CO$_2$ emissions, volcanic activity, solar forcing.
	Stochastic variability  due to temperature uncertainties has been included. The stochastic model for temperature fluctuations is informed by historical
	temperature variability data.
	From top to bottom, $ECS=4.5,3,2.5,2,1.5$}
	\label{fg.everything}
\end{figure}

Figure \ref{fg.forcings} shows the  stochastic long wave and short wave forcing contributions over time. 
The composite indicates  a dominant role for the long wave component during the last 50 years: the trend is larger than its inherent variability.  
In Figure \ref{fg.everything} we depict a single realization of the temperature predictions, driven by the forcing described in Figure \ref{fg.forcings}, {\it i.e.},   the temperature predictions taking into account natural and anthropogenic forcing and their variability.
The upward trend in the long-wave emissivity  still dominates  over any uncertainties due to natural  {\it and} anthropogenic forcings during the Industrial Era.
 In the end  the steady-increasing CO$_2$ forcing is overwhelming the temperature predictions in the Industrial Era and into the future, even when taking into account the variability due to volcanic and solar forcing.


\section*{Summary}
\label{sec:summary}

Without making use of anything more than data we have shown that temperatures around the Northern Hemisphere do not have a time-stationary distribution. More careful analysis of the data has shown that this is a general statement regarding temperature most anywhere on Earth. Fits of the data, for Industrial Era temperatures, show an upward trend. Ocean temperature surveys also indicate that the temperature of the ocean in the upper 2000m has increased. 
The signal-to-noise ratio of this Argo data, collected for the last 20 years, is much higher than near-ground atmospheric temperatures 
\cite{trendb17}. In fact, 
data  and models  indicate that the global mean temperature of the Earth is increasing, since the end of the 19$^{th}$ century
\cite{IPCCAR5}. With a  changing climate, we have observed a changing weather.
  The rate at which climate is  changing is alarming since it is comparable to, or shorter than, the typical relaxation rates of the system (land, ocean, and atmosphere).

An energy balance model, consistent with instrumental data, was used here to explore how the inclusion of inherent uncertainties 
affect  the relative impact of natural and anthropogenic 
forcing on Earth's temperature \cite{sciammann}.  
We estimated uncertainties from the data and constructed  historically-informed models for the variability of each of the  forcings. 
Both of these forcings need to be invoked to get the model to agree with data.  
 Driven by estimated natural and anthropogenic radiative forcing, our calculations indicate that the warming is a result of anthropogenic increases in greenhouse gas concentrations  and that the inclusion of aleatoric uncertainties do not change this outcome in the ensemble sense.   
 Moreover,  since the effect of forcing variability is small compared to the upward trend of the anthropogenic forcing, the inherent variability does not offer much of a reason to expect that the temperature will keep climbing unless there is a  rate slowdown in the anthropogenic forcing, {\it i.e.},  in the production of  GHG. Using far more sophisticated models and state-of-the-art knowledge in climate,  scientists reach the same conclusions \cite{ipcc07} and further, have not been able to find  a non-anthropogenic explanation to the observed increase in warm extremes in  global temperatures  during the Industrial Era ({\it e.g.}, \cite{Mann17}).

 The key natural forcings are associated with volcanic emissions and changes in insolation. 
 While statistical projections of changes in natural volcanic and solar radiative forcing of  climate are by necessity speculative, 
 we find no evidence that such natural radiative forcing changes could substantially alter the projected warming from increasing greenhouse gas concentrations. The impact of their variability would contribute to known unknowns in the temperature uncertainty.
 The long-time features of the model and the historical data agree well and thus
 do not require postulating or requiring epistemic variability (unknown unknowns). 
  
Shah's appraisal of the outcomes in the  Fourth National Assessment report  offered us an invitation to demonstrate how simple, well established, quantitative methods  are used to address apparent  challenges posed by uncertainties in climate assessments. 
Given the evidence that key climate change attributes, such as ice sheet collapse and sea level rise, are occurring ahead of schedule \cite{madhouse} uncertainty has in many respects broken against us, rather than in our favor. Scientific uncertainty is not a reason for inaction. If anything, it is a reason for more concerted efforts to 
limit carbon emissions

\section*{Acknowledgements}
We want to thank Barbara Levi who provided invaluable editorial assistance with this article.

\bibliography{./co2}
\appendix
\section*{Appendix, Stochastic Parameterizations}
We describe the parametrization of the natural and anthropogenic forcing, used in the future projections produced with 
the EBM model, (\ref{ebmeq}). The process for the stochastic parametrization follows the strategy used in \cite{stochlong}.

\begin{figure}[h] \centering
	\includegraphics[scale=0.5,angle=-90]{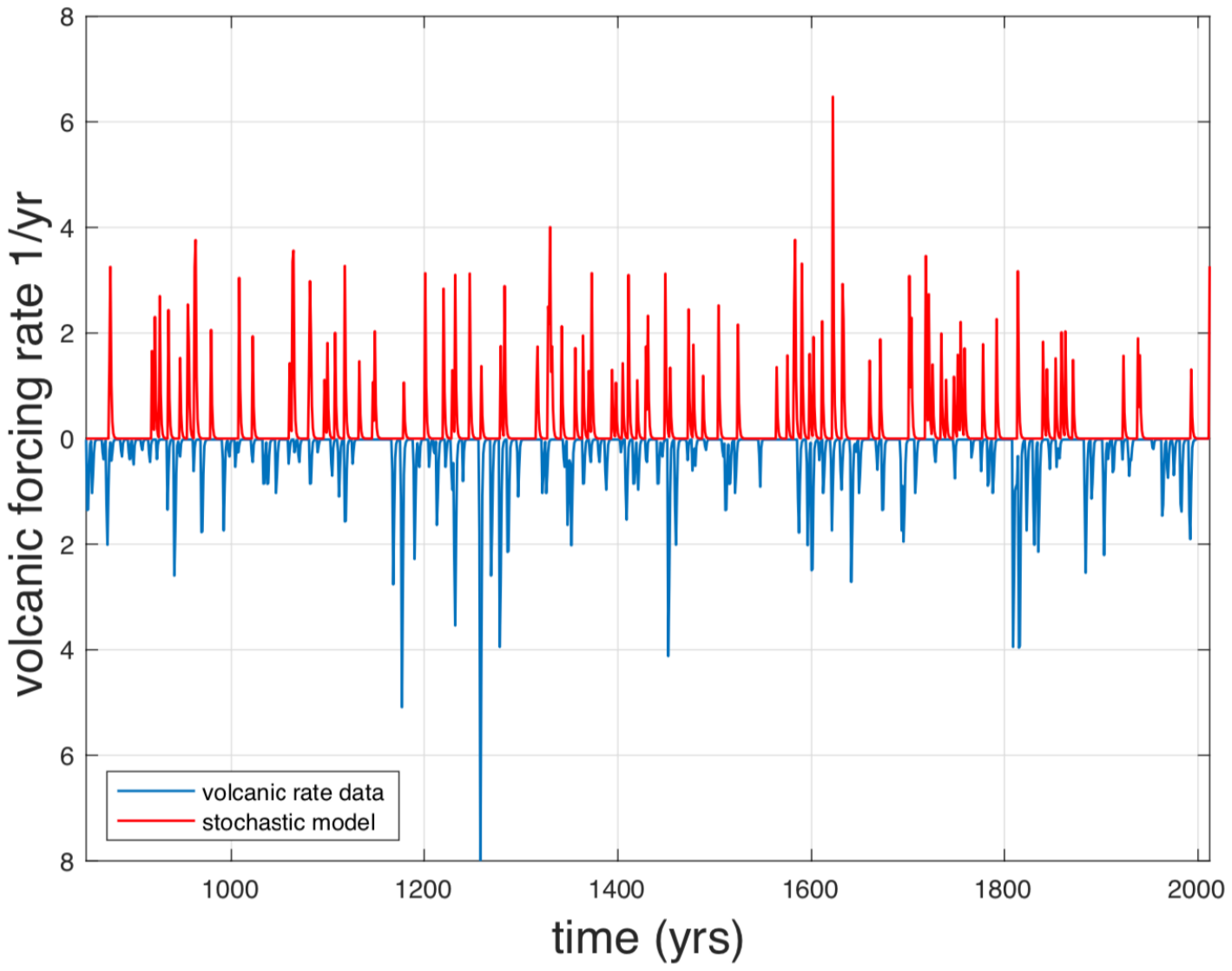}
	\caption{Lower: volcanic forcing data; upper, stochastic parametrization.}
	\label{fg.volc}
\end{figure}

\noindent {\bf Volcanic Emissions:} Using historical data from \cite{ammann}, we propose the following stochastic model:
\[
 dy(t)  = -b  \, y(t) dt+ \exp{({\cal J}_t)}dP_t,
\]
where $b=260/365$, ${\cal J}_t$ is an exponential process with uniform random argument  $1.2  \, {\cal U}[0,1]$, and $dP_t$ is a Poisson incremental  process with $\lambda = 0.07$.
Figure \ref{fg.volc} shows a comparison of the data (lower), and the negative of the stochastic simulation (upper).

\begin{figure}[h] \centering
(a) \includegraphics[height=1.75in, width=4.5in]{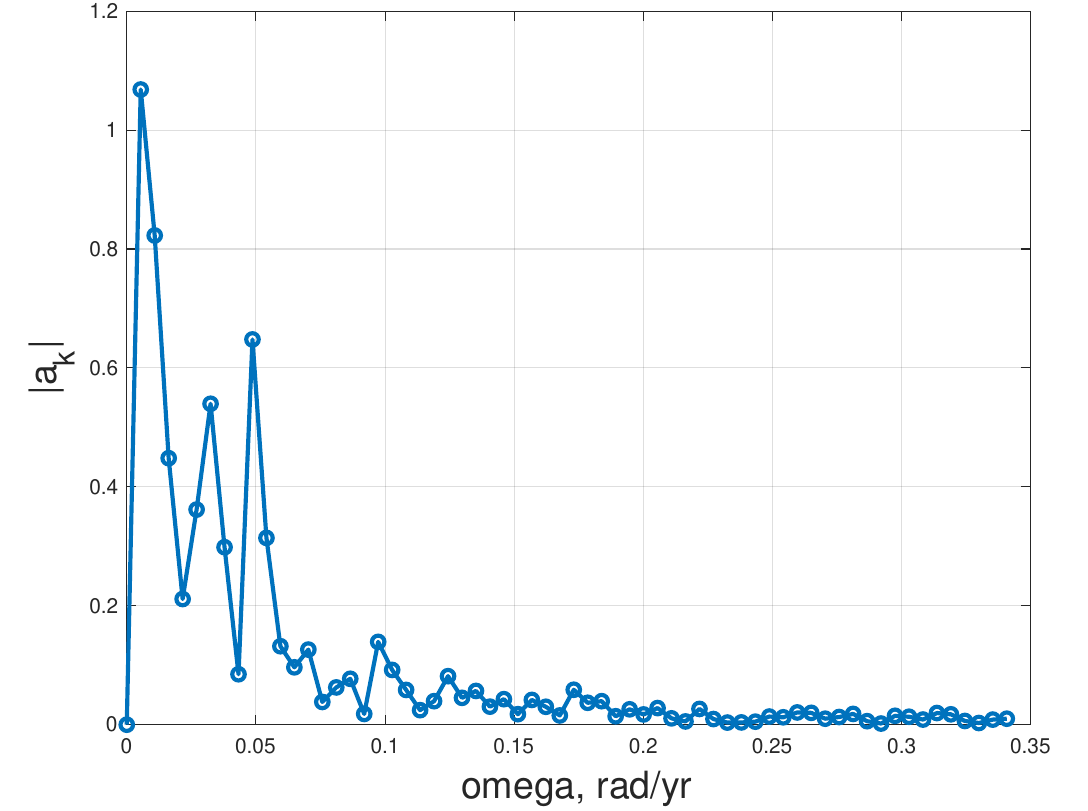}\\
(b) \includegraphics[height=2.5in,width=5in]{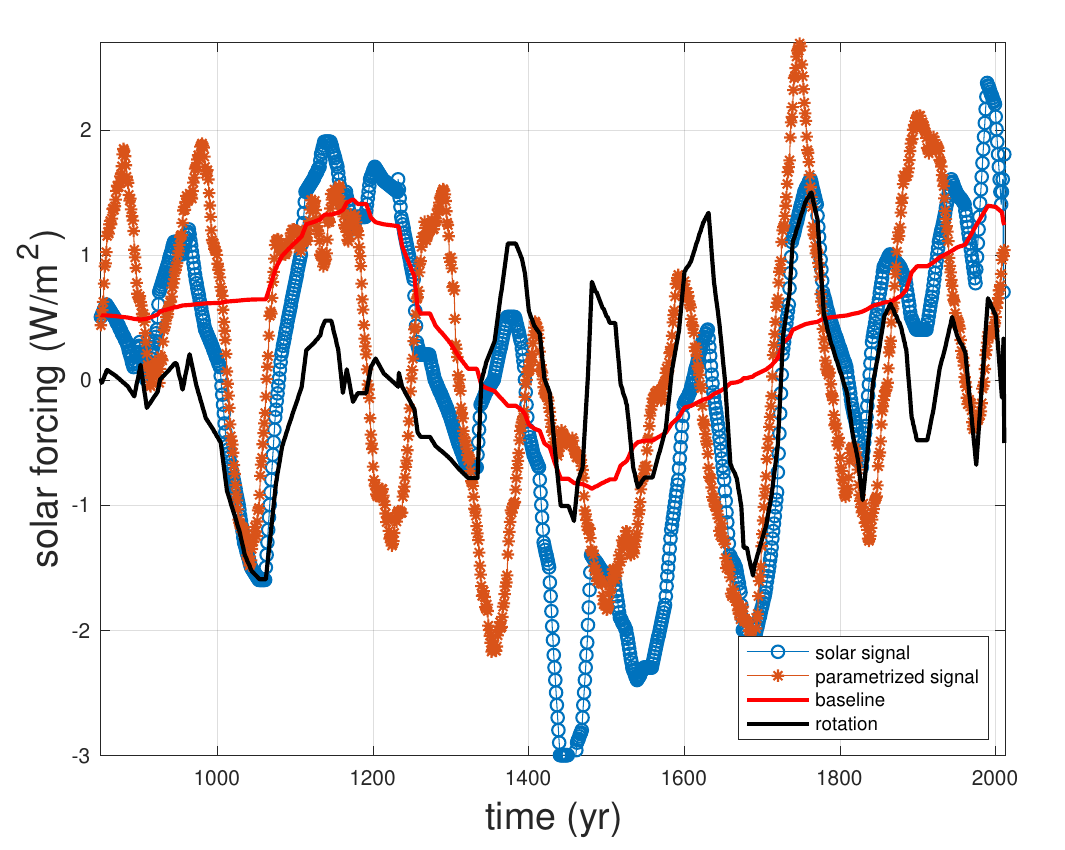}
(c) \includegraphics[height=2.5in,width=5in]{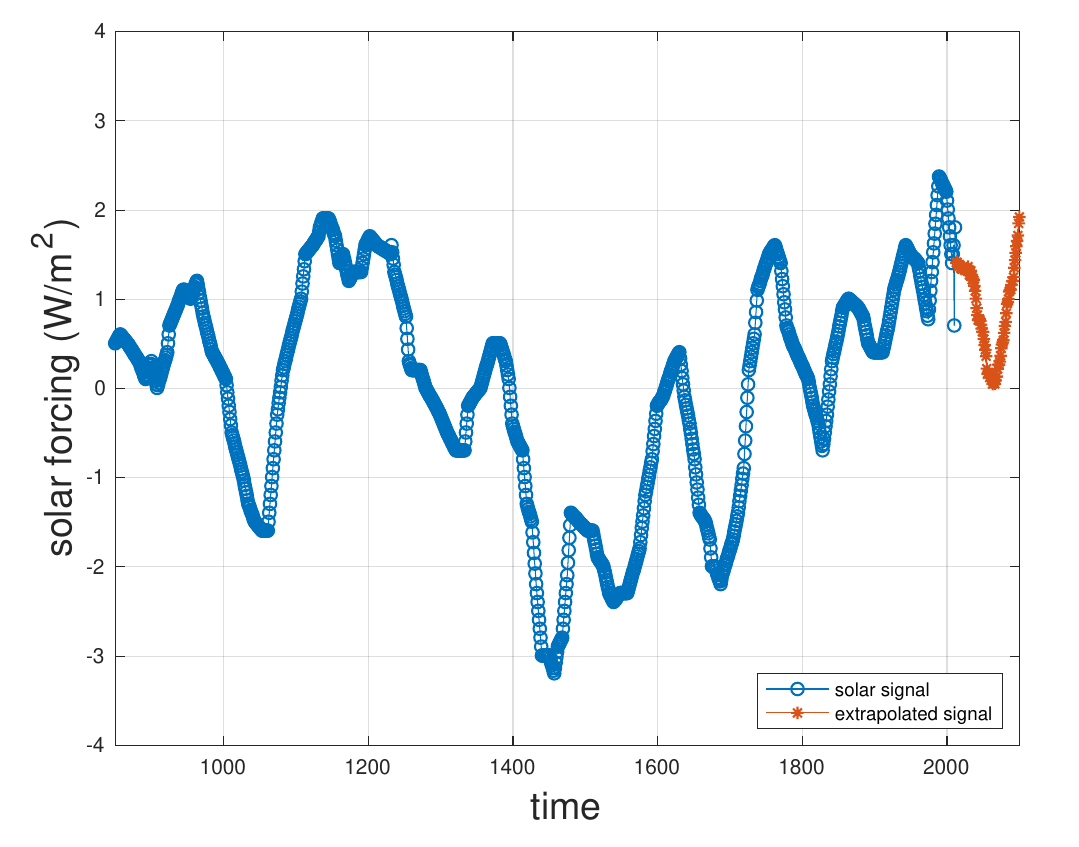}
	\caption{Solar forcing (W/m$^2$). The historical mean 1366 W/m$^2$ is subtracted. (a) Spectrum of the observations; (b) comparison of data and the parametrization
	of the solar forcing. Data (*), parametrization (o). The baseline (blue) and the rotation (red) from the intrinsic time decomposition; (c) observations (o) and extrapolated (*) solar forcing.}
	\label{fg.solarpar}
\end{figure}

\noindent {\bf Solar Forcing:} 
The solar data for the years 850-2012 AD from \cite{ammann} is used.  In what follows, the mean
1365.8 W/m$^2$ is  removed.  Figure \ref{fg.solarpar}a displays the modulus of the spectral 
decomposition of the (periodized) observations. Figure \ref{fg.solarpar}b features the observations and the parametrization, along
with the intrinsic time decomposition of the signal \cite{itdjmr}.
The intrinsic time decompostion of the signal $s(t) = b(t) + r(t)$, is not lossy. The baseline $b(t)$ is designed to lace through the data in such a way that local maximas (minimas) of the time series $s(t)$
are below (above) the baseline at the locations of the extremas. At the extremal locations the baseline is situated at the three-point average of the extreme and its two neighbors. The rest of the baseline is constructed by linear interpolation. The rotation signal is defined by $r(t)=s(t)-b(t)$. The parametrization 
$\sigma(t) = b(t) + n(t)$, where $n(t)$ is an ARMA(4,0)  process obtained from the rotation $r(t)$. 
In the ARMA, the constant was -0.0010598, $p_1=1.7193$, $p_2=-1.192$, $p_3=1.0518$, $p_4=-0.58531$. The standard error on these was less than one percent.
The extrapolation of $\sigma(t)$ beyond the observations, {\it i.e.}, beyond 2012, is obtained by a reset of time in $b(t-t_r)$, where $t_r$ is drawn at random from the years 928 and 1924. The value of the 
solar observations at 2012 is used to adjust the extrapolation. The baseline assumed free boundary conditions at the ends of the observation record. The time scales in the ARMA process was large enough that, for short extrapolations of the signal, in the interval 2012-2100, it was impossible to discern the persistence of the baseline signal. Figure \ref{fg.solarpar}c depicts the signal plus its extrapolation beyond year 2012. 

\begin{figure}[h] \centering
\includegraphics[scale=0.7]{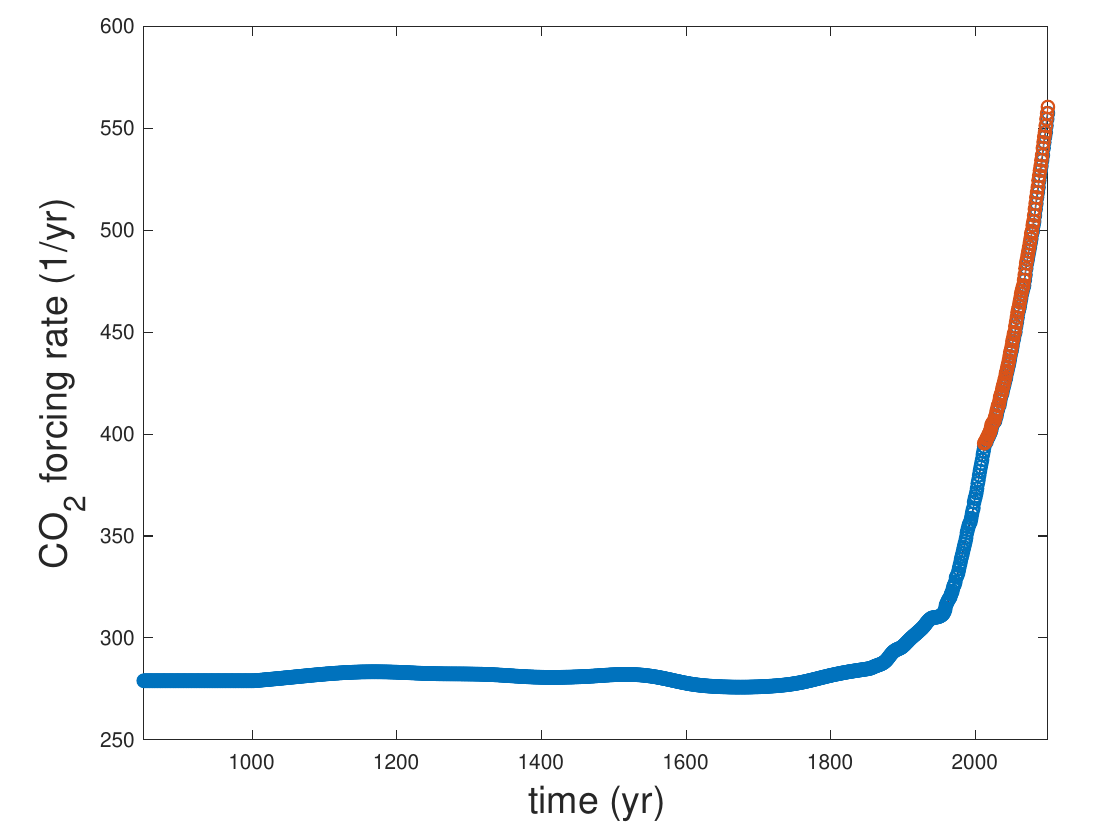}
	\caption{Superposition of the CO$_2$ forcing rate data and the stochastically-fitted 
	version; in red, stochastic parametrization for the years $2012-2100$.}
	\label{fg.co2par}
\end{figure}

\noindent {\bf CO$_2$ Forcing Rate:} 
We use a deterministic fit $\kappa_i(t)$, obtained by least squares, from the CO$_2$ forcing signal. There are two
fitted regions, namely, 
\[
\kappa_1(t)=\sum_{i=0}^4 a_i \tilde y^{i},
\]
where $\tilde y$ = year-1770, with parameters $a_4 = -1.59\times10^{-7}$, $a_3 = 5.93\times 10^{-5}$, $a_2 = -0.006$, $a_1 = 0.265$ and $a_0 = 278$, for the years $1770-1955$, and 
\[
\kappa_2(t) = \sum_{i=0}^2 a_i \tilde y^{i},   
\]
where $\tilde y=$ year-1956, 
for the span $1956-2012$, with parameters $a_2=0.012$, $a_1=0.7943$, and $a_0=394.7128$.
 We combine the variability $T(t) - \kappa_1(t)$ of $1770-1955$, and 
$T(t) - \kappa_1(t)$ of $1956-2012$, and propose an AR(1) model (constant = -0.077584 and $p_1 = 0.80513$), with variance $0.17$ yr$^2$. We then use
$\kappa_2(t)$ to extrapolate to the year 2100 AD, adding the AR(1) process for variability. See Figure \ref{fg.co2par}.

\noindent {\bf Parametrizing Future Temperature Uncertainty:}

The temperature uncertainty $\nu(t)$ is parametrized as follows.
We use  compilations of the Northern Hemisphere temperatures,  for the years 1850-2012 AD (see \cite{sciammann}). 
An ARIMA analysis of the raw data was performed, double differencing the data. The outcomes were satisfactory, but
not as robust as using a fitted removal of the trend and a noise model for the residual.
The parametrization consists of a trend $\tau$ plus a stochastic parametrization of 
$T-\tau$, where $T$ is the normalized temperature record. The fit of the trend is 
\[
\tau=\sum_{i=0}^3 a_i \tilde y^i
\]
where $\tilde y$=year-1850, and 
  $a_3=2.644\times 10^{-7}$,  $a_2=-7.98\times 10^{-6}$, $a_1=-0.000167$, $a_0=-0.274$. The 
  residual $T-\tau$ is represented well (according to the Bayesian Index Criteria, BIC), by an AR(1), 
  with constant $= -3.0333\times 10^{-5}$, $p_1=0.53426$, with errors in the order of 6 percent.
 \begin{figure}[h] \centering
(a) \includegraphics[scale=0.7]{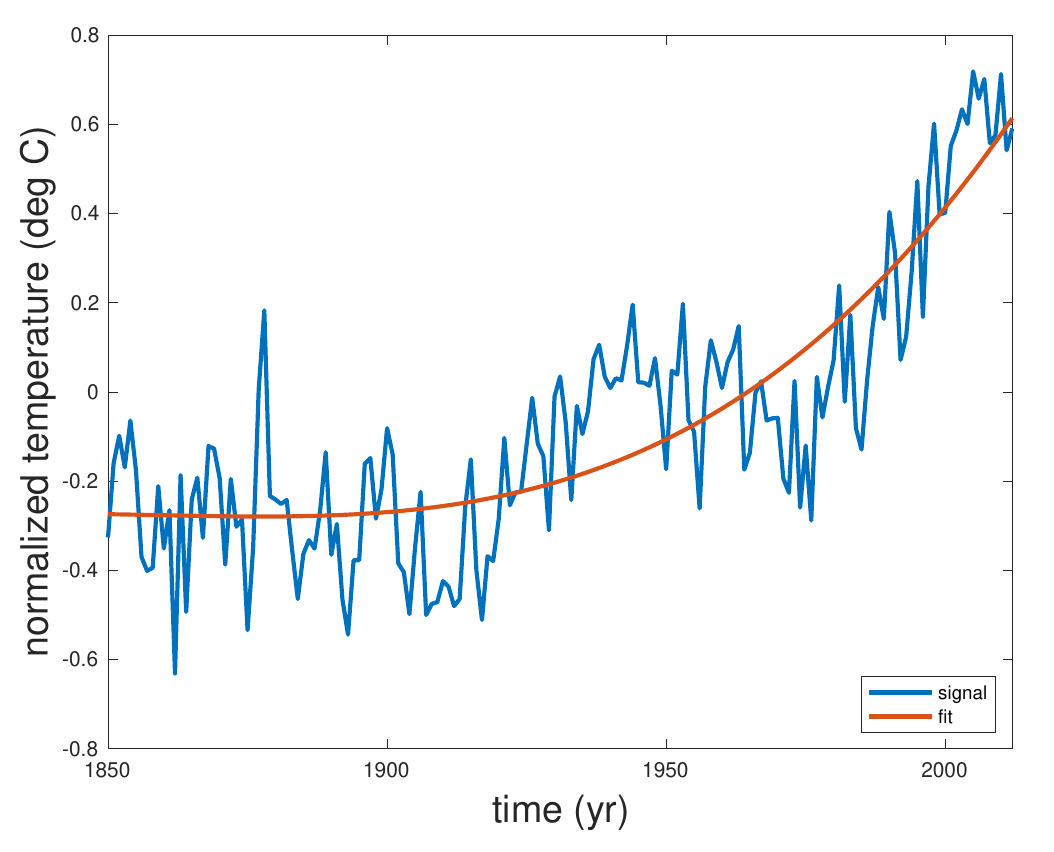}
(b)\includegraphics[scale=0.7]{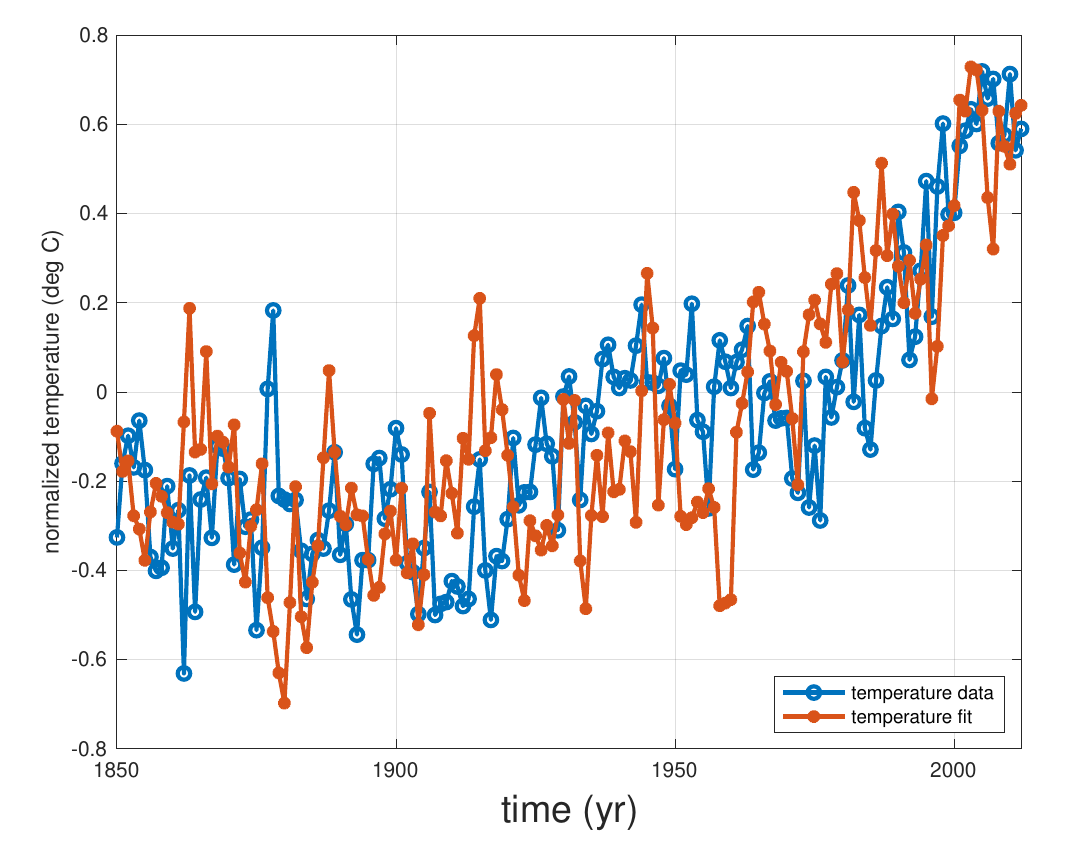}
	\caption{(a) Fit of the trend $\tau=\sum_{i=0}^3 a_i \tilde y^i$, where $\tilde y=$year$-1850$,  and data to the temperature distribution; (b)  stochastic parametrization and data.}
	\label{fg.temppar}
\end{figure} 
\end{document}